\documentclass[aps,reprint,superscriptaddress,amsmath,amssymb,prl,nofootinbib]{revtex4-2}
\usepackage{amssymb}
\usepackage{amsmath}
\usepackage{physics}
\usepackage{latexsym}

\usepackage{verbatim}
\usepackage{mathrsfs}
\usepackage{amsfonts}

\usepackage{graphicx}
\usepackage{subcaption}
\usepackage{epsfig}

\usepackage{units}
\usepackage{overpic}
\usepackage[utf8]{inputenc}
\usepackage{rotating}
\usepackage{enumerate}

\usepackage{soul}

\usepackage{xfrac}
\usepackage{xcolor}
\usepackage{multirow}

\usepackage[hidelinks]{hyperref}
\hypersetup{
    colorlinks,
    linkcolor={red!50!black},
    citecolor={blue!50!black},
    urlcolor={blue!80!black}
}

\usepackage{ragged2e}

\newcommand{\be}{\begin{equation}}

\newcommand{\ee}{\end{equation}}
\begin{document}
\title{Constraints on dark matter cross section with standard model particles from the orbital period decay of binary pulsars}

\author{Gonzalo Agust\'{\i}n Lucero}
\email{glucero@mendoza-conicet.gob.ar} 
\affiliation{Instituto Interdisciplinario de Ciencias B\'asicas, CONICET, Facultad de Ciencias Exactas y Naturales, Universidad Nacional de Cuyo,
Mendoza, Argentina}

\author{Argelia Bernal}
\affiliation{Departamento de F\'isica, Divisi\'on de Ciencias e Ingenier\'ias \\
Campus Le\'on, Universidad de Guanajuato,
Le\'on 37150, M\'exico}

\author{Juan Barranco}
\affiliation{Departamento de F\'isica, Divisi\'on de Ciencias e Ingenier\'ias \\
Campus Le\'on, Universidad de Guanajuato,
Le\'on 37150, M\'exico}

\author{Andr\'es Ace\~na}
\affiliation{Instituto Interdisciplinario de Ciencias B\'asicas, CONICET, Facultad de Ciencias Exactas y Naturales, Universidad Nacional de Cuyo,
Mendoza, Argentina}
\vspace*{1cm}
\begin{abstract}
It is shown that the data from the orbital period decay of binary pulsars give strong constraints on the dark matter-nucleons cross section. The limits are robust and competitive because this new method for testing dark matter interactions with standard model particles has a minimal number of assumptions combined with the extremely high accuracy on the measurement of the decay rate of the orbital period of binary systems. Our results exclude (with $95\%$ confidence) spin independent interactions with cross sections greater than $3.1 \times 10^{-31}\,\mbox{cm}^2$ for a dark matter particle with mass $m_\chi c^2=1\,\mbox{keV}$ and $3.7 \times 10^{-31}\,\mbox{cm}^2$ for $m_\chi c^2=1\,\mbox{TeV}$ which improves by several orders of magnitude previous constraints. 
\end{abstract}
\maketitle

%
A large number of theoretically well motivated dark matter (DM) models suggest that DM could interact directly with particles in the standard model (SM) \cite{Jungman:1995df,Feng:2010gw}. 
Evidence of this non-gravitational interaction may reveal the DM fundamental nature. Substantial efforts to detect DM-SM interactions trough direct detection experiments \cite{PandaX-4T:2021bab,LZ:2022lsv,XENON:2023cxc} or indirect methods \cite{Barman:2022jdg,Chan:2019ptd,Song:2023xdk} have constrained or will constrain interacting DM models. 
Most of these experiments are optimized for DM particles with masses in the range between $1-100\,\mbox{GeV}$.  The sensitivity of direct detection experiments rapidly drops for masses lower than $1\,\mbox{GeV}$ because the nuclear recoil energy is below the detector energy threshold.  On the other hand, for masses higher than $1\,\mbox{TeV}$, direct detection sensitivity decreases because the expected fluxes become tiny in most of the terrestrial detectors.     

Motivated by the necessity to prove DM-SM interactions both in the low particle mass regime (below the MeV) and in the high mass regime, it is imperative to look for other physical processes. In particular, for light DM, the strongest constraints of interactions with nucleons come from cosmic microwave spectral distortions \cite{Ali-Haimoud:2015pwa,Gluscevic:2017ywp}. For heavy non-annihilating DM, the strongest constraints come from 
the observation of old neutron stars that have survived the possible collapse into black holes induced by the capture of DM \cite{Goldman:1989nd,Gould:1989gw,Bertone:2007ae,deLavallaz:2010wp,McDermott:2011jp,Kouvaris:2010jy,Bramante:2014zca,Bhattacharya_2023}.

In this work, we show that the orbital period decay rate $\dot{P_b}$ of millisecond binary pulsars (MBP) can constrain the dark matter cross section with standard model particles in both light and high mass regimes. 

The idea is simple enough: 
if DM particles scatter with nucleons in a pulsar, they could lose sufficient energy to be captured by the pulsar's gravitational field. If we consider that DM does not self-annihilate, it will simply accumulate and increase the mass of the pulsar. If the capture process is very efficient, the change in mass of the pulsars will affect $\dot{P_b}$, which is one of the most precise physical quantities ever measured.  Surprisingly, this possibility has not been considered before. 
As it is known, due to the short spin period  and the high stability  of a binary pulsar system,
it is possible to have an extremely accurate monitoring of the times of arrival of the pulsar's radio  pulses 
from which the Keplerian and Post-Kleperian (PK) parameters can be inferred with high accuracy. One of the PK parameters is $\dot{P}_b$. In general relativity (GR), the PK parameters depend only on the two masses of the pulsars, which have been determined with a precision of one part in a ten-thousand. Thus, binary pulsars have been used as useful tools to test GR. We take advantage of this high precision in the determination of the mass of the pulsars 
in order to place constraints on the DM-SM cross section.


First, we need to consider the capture of DM by neutron stars (i.e. pulsars). This process has been studied for a long time now \cite{Goldman:1989nd,Gould:1989gw,Bertone:2007ae,deLavallaz:2010wp,McDermott:2011jp,Kouvaris:2010jy,Bramante:2013nma,Bramante:2014zca,Bhattacharya_2023}. For completeness, here we will consider the capture rate given by \cite{Bhattacharya_2023}
\begin{table*}
\begin{center}
\begin{tabular}{|c | c | c | c | c | c | c | c | c | }
\hline $i$ &Name & $P_b$ (days) & $\dot{P}_b^{exp}$ & $e$ & $m_{c}$ & $m_{p}$ & Reference\\ \hline
$1$ & J0737-3039 S & $0.10225156248(5)$ & $-1.252(17)\times10^{-12}$ & $0.0877775(9)$ & $1.2489(7)$ & $1.3381(7)$ &\cite{Kramer:2006nb}  \\ \hline
$2$ &J1756-2251  & $0.31963390143(3)$ & $-0.234^{+0.009}_{-0.006}\times10^{-12}$ & $0.1805694(2)$ & $1.230(7)$  & $1.341(7)$   & \cite{Fonseca:2014qla,Stairs:2002cw}\\ \hline
$3$ & B1913+16 & $0.322997448918(3)$ & $-2.398(4)\times10^{-12}$ & $0.6171340(4)$ & $1.390(1)$ & $1.438(1)$ &\cite{Weisberg:2016jye}\\ \hline
$4$ & J1757-1854 & $0.18353783587(5)$ & $-5.3(2)\times10^{-12}$ & $0.6058142(10)$ & $1.3946(9)$ & $1.3384(9)$ &\cite{Cameron:2017ody}  \\ \hline
\end{tabular}
\caption{Measured values of the orbital parameters with their experimental error for binary pulsar systems.} \label{measured}
\end{center}
\end{table*} 
\begin{widetext}
\begin{equation}
    C_{NS} = 1.4\times 10^{20}\,\mbox{s}^{-1}\left(\frac{\rho_{\chi}}{0.4\,\mbox{GeV cm}^{-3}}\right)\left(\frac{10^{5}\mbox{GeV}}{m_{\chi}}\right)\left(\frac{\sigma_{\chi}}{10^{-45}\,\mbox{cm}^{-2}}\right)
     \left[1-\frac{1-e^{-A^{2}}}{A^{2}}\right]\left(\frac{v_{esc}}{1.9\times 10^{5}\,\mbox{km s}^{-1}}\right)^{2}\left(\frac{220\,\mbox{km s}^{-1}}{\bar{v}_{gal}}\right),\label{capture}
\end{equation}
\end{widetext}
where $\rho_\chi$ is the DM density, $m_\chi$ the mass of the DM particle, $v_{esc}$ is the neutron star escape velocity, and $\bar{v}_{gal}$ is the DM speed in the galaxy and $\sigma_\chi$ is the DM-SM cross section, which we consider independent of the relative particle velocity as a first order estimate. Elaborated analysis where broader DM-SM interactions with wider momentum and velocity dependent interactions can be incorporated in the future. 
\begin{figure*}
    \centering
    \includegraphics[scale=0.23]{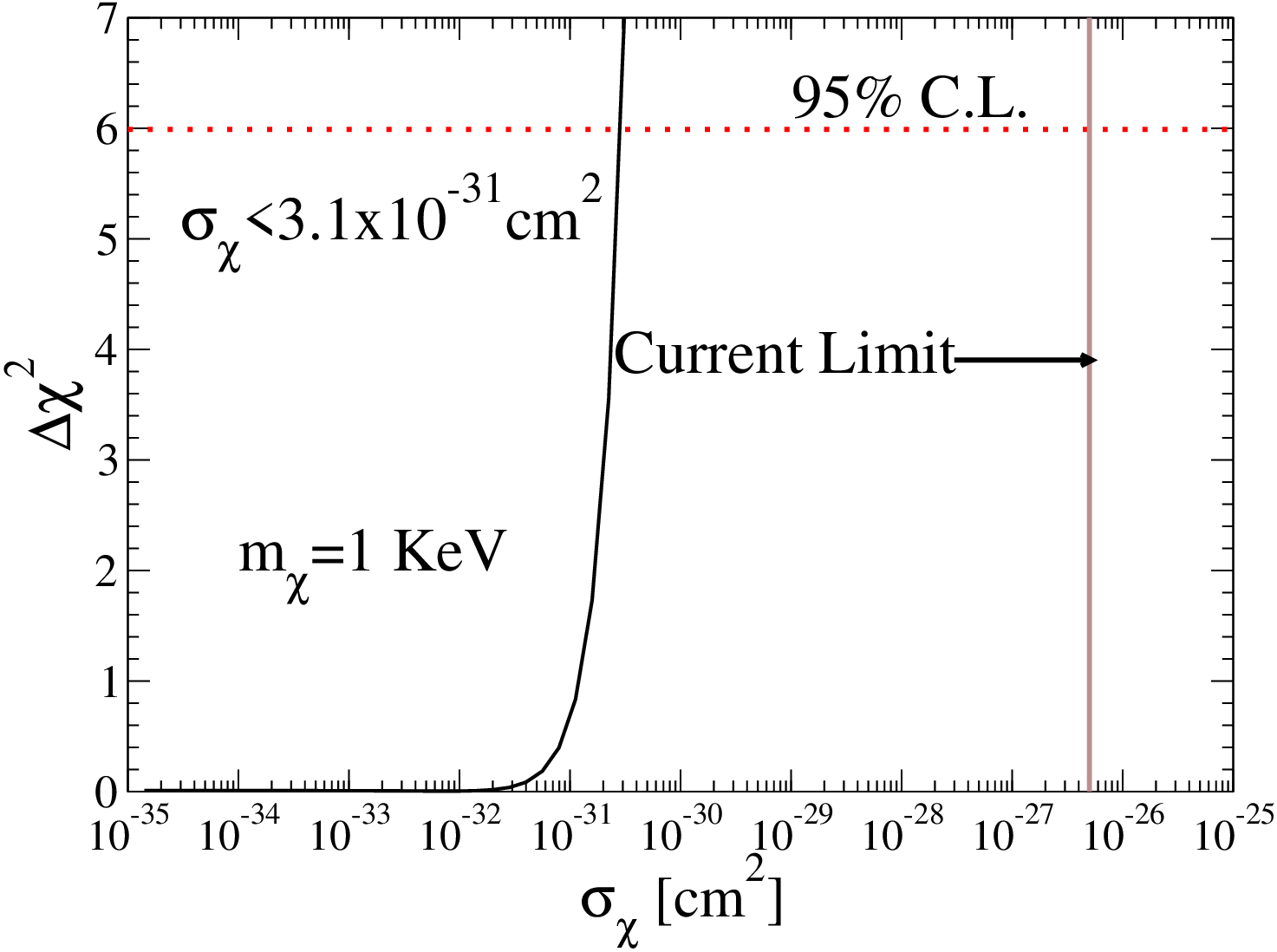}
   \includegraphics[scale=0.23]{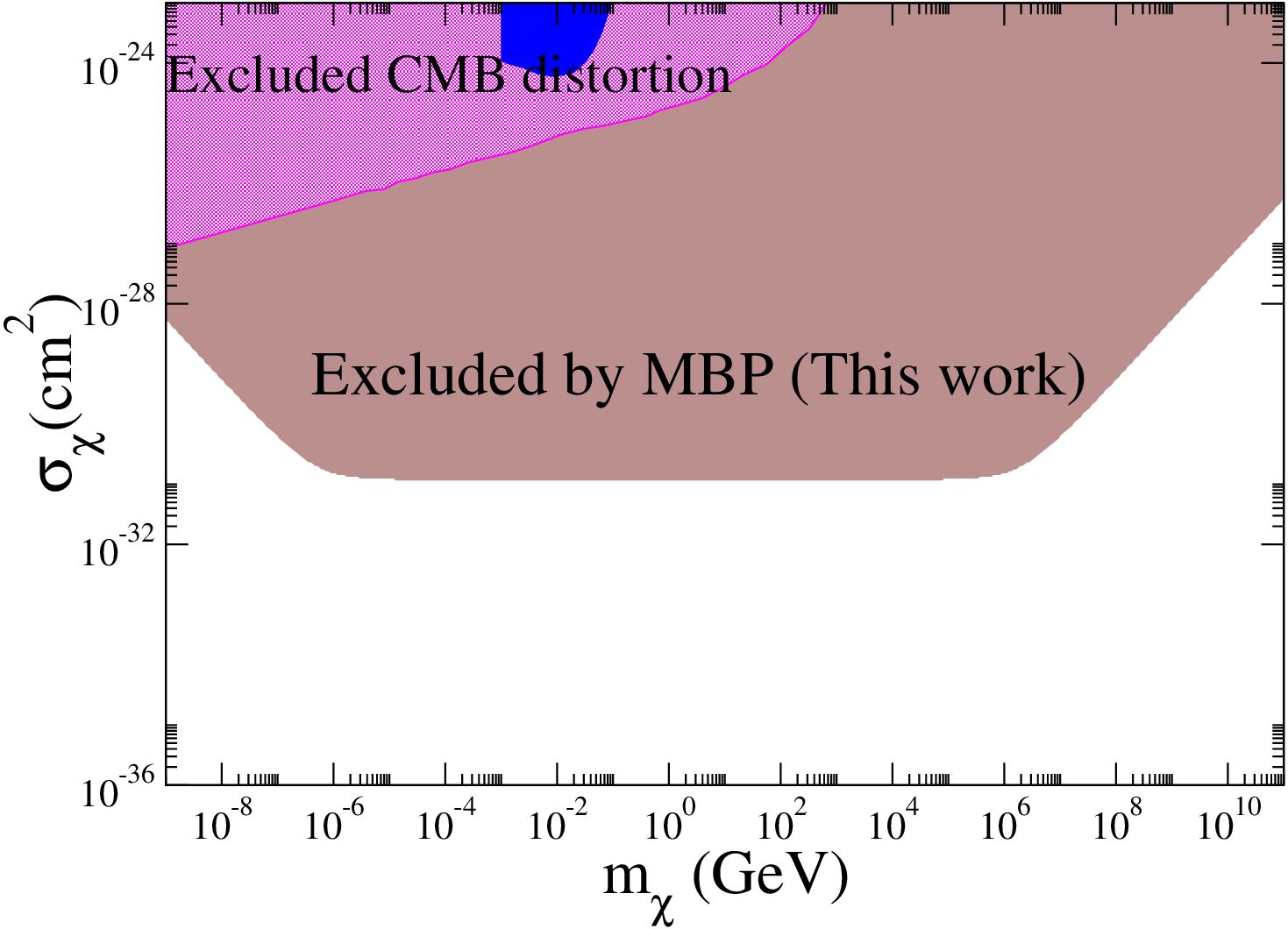}
   \includegraphics[scale=0.23]{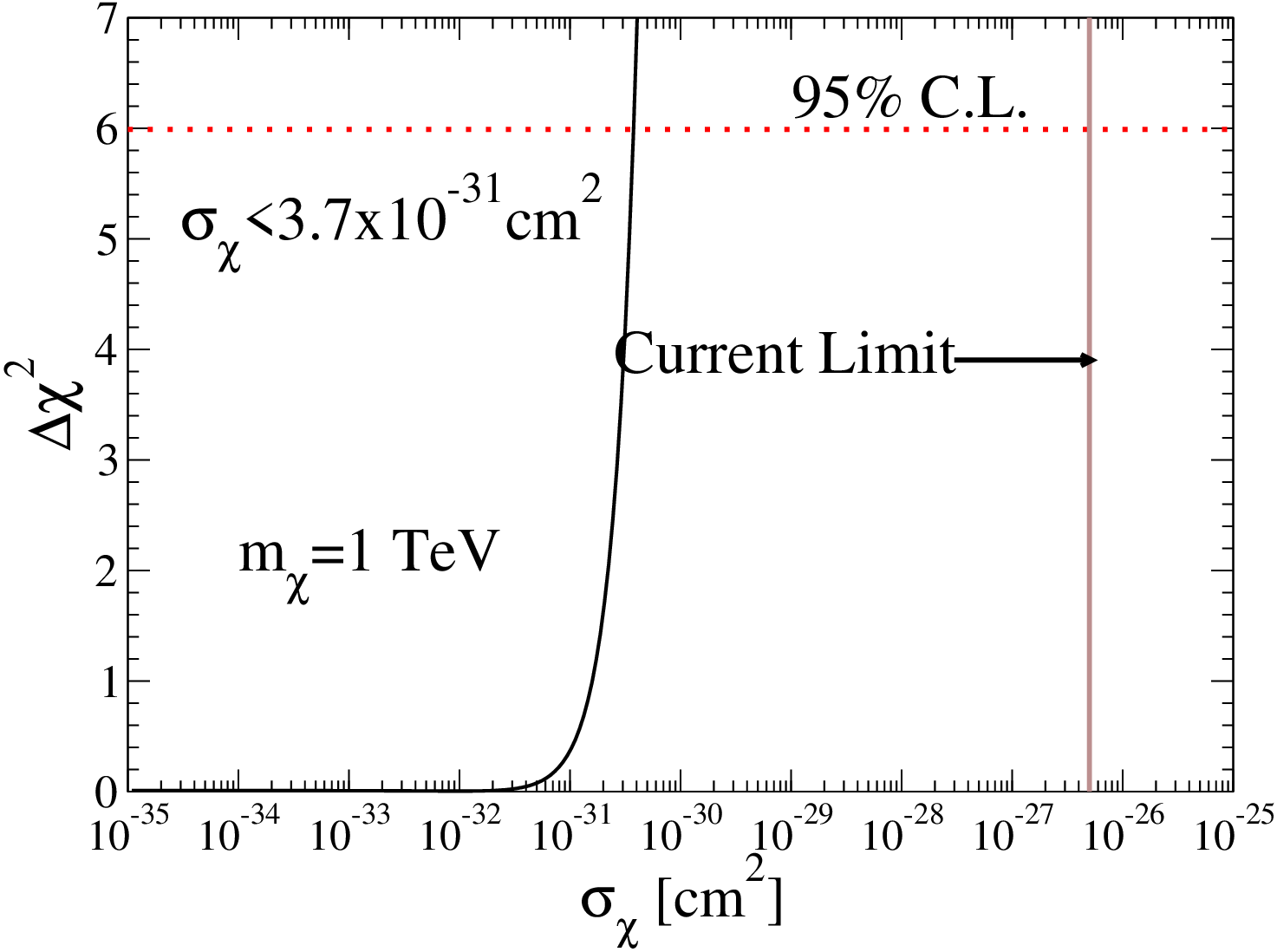}
    \caption{Middle panel shows the constraints on the $(m_\chi,\sigma_{\chi N})$ space at $95\%$ C.L. obtained with the decay of the orbital period of millisecond binary pulsars (MPB). For references, limits obtained for light DM candidates are also plotted \cite{Ali-Haimoud:2015pwa, Gluscevic:2017ywp}.  Left and right panel shows cuts in $\Delta \chi^2$ function for fixed values of $m_\chi=1$KeV (left panel) and $m_\chi=1$ TeV (right panel).   }
    \label{chi2vssygma}
\end{figure*}
Additionally, in Eq. \ref{capture}, the term in square brackets incorporates the effect of DM particles that are too energetic to be captured by the pulsar and the term $A$ is given by \cite{Bramante:2013nma,Bramante:2014zca,Bhattacharya:2023stq} 
\begin{equation}
A^2=\frac{v_{esc}^2 m_\chi m_N}{\bar{v}(m_\chi-m_N)^2}\,,
\end{equation}
where $m_N$ is the mass of a SM nucleon.
The capture rate of DM particles by a pulsar can be extremely high, depending on $\sigma_\chi$. For instance, from Eq. \ref{capture}, if $\sigma_\chi=10^{-20}\,\mbox{cm}^2$, in one year, a pulsar can increase its mass by $\Delta M\sim 0.1M_\odot$ if the DM's particle mass is $m_\chi=1\,\mbox{GeV}$. Thus, a system where it is possible to follow the mass evolution of a neutron star over a period of years could set limits on the DM-nucleon cross section. 
This is precisely the case of microsecond pulsar binaries, where a change in the mass of the pulsars might affect the orbital decay rate $\dot{P_b}$. 

Indeed, let us consider a binary pulsar system, with masses $m_p$ and $m_c$, in a Keplerian orbit. The period orbit  can be computed as
\be\label{P0}
P_b = \frac{2\pi GM}{(-2E)^\frac{3}{2}}\,,
\ee
where we have used Kepler's third law expressed in terms of the Newtonian energy of the system, $E = -\frac{GM}{2a}$, and $M=m_p+m_c$ is the total mass of the binary system.
Thus, the change in the period with respect to time, considering that $M$ can change, but it is independent of $E$, will be simply:
\be
\frac{\dot{P_b}}{P_b} = -\frac{3}{2}\frac{\dot{E}}{E} + \frac{\dot{M}}{M}. \label{dotP}
\ee
In GR, the orbital period decreases due to the emission of gravitational waves. The average rate of energy emission in the post-Newtonian expansion to order zero is given by
\cite{Peters:1963ux,Peters:1964zz,Blanchet:1989cri}
\begin{equation}
\dot{E} = -\frac{32G^4}{5c^5}\frac{\mu M^3}{a^5(1-e^2)^\frac{7}{2}}\left(1+\frac{73}{24}e^2+\frac{37}{96}e^4+...\right), \label{dotE}
\end{equation}
where $\mu$ is the reduced mass of the system, $a$ the semi-major axis and $e$ the eccentricity of the ellipse. 
$\dot{P_b}$ has been measured with extremely high precision. In Table \ref{measured} we report the orbital parameters measured for four binary pulsar systems \cite{Kramer:2006nb,Fonseca:2014qla,Weisberg:2016jye,Cameron:2017ody}.
It is worth noticing the Hulse-Taylor binary pulsar PSR B1913+16 \cite{Hulse:1974eb}, first observed in the 70s, which became the first indirect detection of gravitational waves since the prediction of the decay in $P_b$, is consistent with Eq. \ref{dotP} and Eq. \ref{dotE} for $\dot{M}=0$.  

Furthermore, if we consider that both neutron stars in the binary accrete DM, we can use the relationship between the accretion rate and the variation in the total mass of the system:
\begin{equation}
    \frac{\dot{M}}{M}=\frac{(C_{m_p}+C_{m_c})}{m_{p}+m_{c}}m_{\chi}.
\end{equation}
Thus, we have a connection between DM models and the
measured quantity $\dot{P_b}$. Indeed, the theoretical decay rate will be given by
\begin{widetext}
 \begin{equation}
   \frac{\dot{P}_b}{P_b}(m_\chi,\sigma_\chi)=-\frac{(2\pi G T_{\odot})^{5/3}}{P_b^{8/3}}\frac{2 \pi m_p m_{c}}{M^{1/3}(1-e^{2})^{7/2}}\Big\lbrace{\frac{96}{5}(1+\frac{73}{24}e^{2}+\frac{37}{96}e^{4})+...\Big\rbrace}+\frac{(C_{m_p}(m_\chi,\sigma_\chi)+C_{m_c}(m_\chi,\sigma_\chi))}{m_{p}+m_{c}}m_{\chi},
\end{equation}   
\end{widetext}
with $T_{\odot}\equiv GM_{\odot}=4.925\,\mu\mbox{s}$, and $C_{m_p}$ and $C_{m_c}$ are given by Eq. \ref{capture} for the neutron stars of the binary system with masses $m_c$ and $m_p$.

In order to obtain constraints on the DM particles we perform a $\Delta \chi^2$ analysis based on the free parameters, which in our case were the DM-SM cross section $\sigma_{\chi}$, and the mass of the DM particle $m_\chi$,
\begin{equation}
\chi^2(m_\chi,\sigma_\chi)=\sum_{i=1}^4\left(\frac{\frac{\dot{P_b}}{P_b}(m_\chi,\sigma_\chi)-\frac{\dot{P_b}}{P_b}^{exp}_i}{\delta\frac{\dot{P_b}}{P_b}_i}\right)^2 \,,   
\end{equation}
where $i$ runs for the data taken from Table \ref{measured}.
This allows us to obtain confidence intervals that are shown in the middle panel of
Fig. \ref{chi2vssygma}, where we have plotted iso-curves for $\Delta \chi^2(m_\chi,\sigma_\chi)=\chi^2(m_\chi,\sigma_\chi)-\chi^2_{min}$ at $95\%$ C.L. For comparison, similar bounds obtained in the same parameter space, $(m_\chi,\sigma_\chi)$, obtained previously by analysis of spectral distortion on the cosmic background microwave \cite{Ali-Haimoud:2015pwa,Gluscevic:2017ywp}, are shown. It is clear that the orbital period decay of binary pulsars strongly constrains the DM-SM cross section. 
%
Additionally, we have plotted in Fig. \ref{chi2vssygma} the cuts in the $\chi^2$ function for two representative values of $m_\chi$. The first panel corresponds to $\chi^2(m_\chi=\mbox{1 KeV},\sigma_\chi)$ and the third panel to $\chi^2(m_\chi=\mbox{1 TeV},\sigma_\chi)$. We can see that our method improves previous constraints for those masses by several orders of magnitude. Namely, we exclude (with $95\%$ confidence)  cross sections greater than $3.1 \times 10^{-31}$cm$^2$ for a DM particle with mass $m_\chi c^2=1$ keV and $3.7 \times 10^{-31}$ cm$^2$ for $m_\chi c^2=1$ TeV that improve previous limits found in \cite{Ali-Haimoud:2015pwa,Gluscevic:2017ywp}.  


%

\section*{Acknowledgments}
This work was partially supported by CONAHCYT-SNII and CONAHCYT Projects CF-2019/304001 and CF-2019/376127 and by Agencia I+D+i (Argentina) through grant PICT-2021-I-INVI-00597. 
\bibliographystyle{unsrt}
\bibliography{pulsars.bib} 
\end{document}